\documentclass[epj,referee]{svjour}
\usepackage{graphics}
\begin{document}
\title{Ground state correlations in Deep Inelastic Scattering and the Drell-Yan process.}
\author{O.~Linnyk 
        \and S.~Leupold 
        \and U.~Mosel 
}
\institute{Institut fuer Theoretische Physik, Universitaet Giessen, Heinrich-Buff-Ring, 16, D-35392 Giessen, Germany}
\date{Received: date / Revised version: date}
%
\abstract{ Such high energy processes as deep inelastic scattering
(DIS) and Drell-Yan heavy lepton pair production are among the
most important tools to probe the quark and gluon interaction. We
calculated non-perturbative corrections to the LO cross section
formulae for DIS and the Drell-Yan process.  The interaction of
partons was taken into account via dressing the incoming quark
lines with spectral functions. We
found that the effect of ground state correlation in DIS is large
in the region of small Bjorken $x$ and low momentum transfer
$Q^2$.  For a quark width of the order of 200~MeV, the cross
section deviation reaches as much as 50\% for $Q^2=10$~GeV$^2$. On
the other hand, for the values of $Q^2$ well above the resonance
region, e. g.  200~GeV$^2$, the effect of the initial quark
off-shellness turned out to be small in DIS, but still substantial for
the triple differential Drell-Yan cross section.
Semi-inclusiveness of the latter process opens a possibility to
extract from the experimental data important information on the shape of
the quark spectral function in the nucleon. From the 
comparison 
to the
resent data on the Drell-Yan cross section from NuSea
collaboration, we obtained for the width of the quark spectral
function the value about 200~MeV. The performed off-shell DIS and
Drell-Yan cross section calculations allow for a better
understanding of the quark and gluon interaction in the nucleon
and, thus, the nucleon structure.
} 
\maketitle
\section{Introduction}

\label{intro}

One of the major goals of present day research is to study the
structure of the nucleon and other hadrons in terms of the
fundamental quark-gluon dynamics. One would like to gain as good
an understanding of hadron structure as our understanding of the
compositeness of the nucleus in terms of nucleons and their
interaction.

In high energy hadronic
processes like deep inelastic scattering (DIS), the Drell-Yan
process, jet production, etc. the soft and hard subprocesses can
be disentangled. This allows one to apply the well established
methods of perturbative QCD for the extraction of the information
about the non-perturbative quark and gluon properties in a bound
state from the experimental data. This way, for instance, the
distributions of partons, having different hadron light cone
momentum fractions (parton distribution functions), are found 
\cite{grv}. 
The
 
described 
method, 
based on 
factorization
principle, 
is
analogous to plane wave impulse approximation (PWIA) for the
quasi-elastic (e,e'p) scattering in nuclear physics. In the theory
of nuclei the importance of the effects beyond the PWIA, e.q.
photon radiation, initial state interaction (ISI), and final state
interaction (FSI) is well understood. 
Semi-\-exclu\-sive 
processes
offer an opportunity to study these effects. Cross section
measurements, in which energy and momentum of the nucleon can be
determined from the final state kinematics, can thus probe nuclear
structure via the spectral function.

The neglected in PWIA initial and final 
state quark interaction
effects on the observable hard scattering 
cross sections has not
been studied yet.
 
The primordial transverse 
momentum of the quarks
in the nucleon has been
 
considered in some 
works~\cite{kt}, 
but the effect of the parton
 
virtuality on the observed 
inclusive and
semi-exclusive processes
 
has not been calculated 
so far. 

As we show below, the
off-shellness effects 
have the same order of magnitude as those of
the intrinsic transverse
momentum. Thus, the consistent treatment of the both off-shellness
and 
non-collinearity 
is necessary. By properly taking into account the
ISI and off-shell kinematics of quarks, we succeeded to 
simultaneously describe
the experimentally measured 
fully inclusive
 
(for example, DIS) as 
well as semi-inclusive (triple-differential
cross section of the Drell-Yan process) cross sections very well.

The applied technique and obtained results are presented in the following
sections.

\section{Method}

\label{method}

The basic tool in the calculation of hard processes is the
factorization into hard and soft physics: $$ d \sigma = f (Q^2,
\xi ) \otimes d \hat \sigma (\xi). $$ We additionally 
took into account the
initial state interaction via dressing the incoming quark lines
with a spectral function and used the generalized
factorization: $$ d \sigma = f (Q^2 , p_T , \xi ) \otimes d \hat
\sigma (\xi , m) \otimes Sp (m,\Gamma) , $$ where: $d \hat \sigma
(\xi , m)$ - off-shell  partonic cross section, $Sp (m,\Gamma)$ -
quark spectral function(s), $f (Q^2 , p_T , \xi )$ - unintegrated
quark distribution, $\xi$ - hadron light cone momentum fraction carried
by the struck parton, $m$ - quark virtuality, $\Gamma$ - quark width.

The hard part, i.e., partonic cross section is calculated using
the rules of perturbative quantum chromodynamics (pQCD). We have
calculated the pQCD cross section of electron scattering off a
virtual quark and the cross section of the annihilation of an
off-shell quark-antiquark pair into a pair of dileptons.
Both off-shell cross sections turn out
to be gauge invariant. So, the modification of the vertex by the
Ward identity was not necessary.

The full kinematics was taken into account as well. In case of DIS it reads:
$$
\xi=\frac{x}{Q^2}\left(Q^2-2\vec{k}_\perp\cdot\vec{q}_\perp-m_i^2\right),
$$

Thus, the struck quark's virtuality and the hadron light cone
momentum fraction, carried by this quark, are linearly connected.
This is not the case in the Drell-Yan process. For the kinematics
of Drell-Yan process in the hadron center of mass system, we
obtained the following relations:
\begin{eqnarray}
\nonumber
M_{DY}^2=m_1^2+m_2^2 + \xi_1 \xi_2 P_1 ^+ P_2 ^-
+ \frac{\left( m_1^2 +\vec{p}_{1\perp}^2 \right)
\left( m_2^2 +\vec{p}_{2\perp}^2 \right)}{\xi_1\xi_2 P_1^+ P_2^-}
\\
\nonumber
x_F = \frac{1}{\sqrt{S}}
\left(
\xi_1 P_1 ^+ -\xi_2 P_2 ^-
- \frac{\left( m_1^2 +\vec{p}_{1\perp}^2 \right)}{\xi_1 P_1 ^+}
+ \frac{\left( m_2^2 +\vec{p}_{2\perp}^2 \right)}{\xi_1 P_2 ^-}
\right)
\end{eqnarray}

In the target rest frame, the connection between the observable mass
($M_{DY}^2$), Feynman variable ($x_F$), and transverse
momentum ($p_T$) of the lepton pair on one hand side and the
partonic variables on the other hand side is
simpler. However, factorization in the usual form is not applicable in this 
system of
 
reference \cite{us}.

The following parameterization is commonly used for the
unintegrated parton distributions: $$f (Q^2 , p_T , \xi )=f(p_T)
\cdot q (Q^2,\xi),$$ where $f(p_T)$ - a Gaussian, $q (Q^2,\xi)$ -
conventional parton distribution functions. For the latter, we used
the
latest parameterizations by Glueck, Reya, Vogt~\cite{grv}.

In our calculations, a Breit-Wigner parameterizations for the
quark spectral function was applied. The width was considered
constant for a constant hard scale of the process ($Q^2$ for DIS
and $M_{DY}^2+p_T^2$ for Drell-Yan).

\section{Results}

\label{results}

\begin{figure}
  \includegraphics{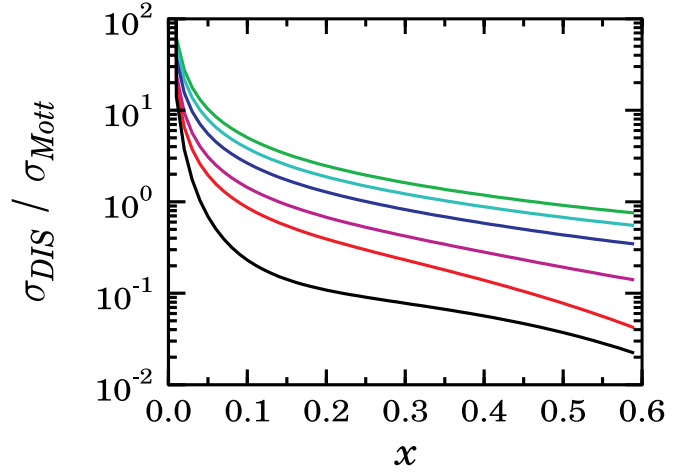}
\caption{DIS cross section for the range of quark spectral
function widths from 5~MeV to 1~GeV and the parton model (the
lowest line).} \label{dis1}
\end{figure}

The result of our calculations for DIS for a range of widths as
compared to the parton model is shown in figure~\ref{dis1}. We
found that the effect of the initial state interaction in DIS is
large in the region of small Bjorken x and low momentum
transfer $Q^2$.  For the quark width 200~MeV, the cross section
deviation reaches as much as 50\% for $Q^2=10$~GeV$^2$. On the
other hand, for the values of $Q^2$ well above the resonance
region, e. g.  200~GeV$^2$, the effect of the initial quark
off-shellness accounts only to at most 10\% of the LO cross
section.

\begin{figure}
  \includegraphics{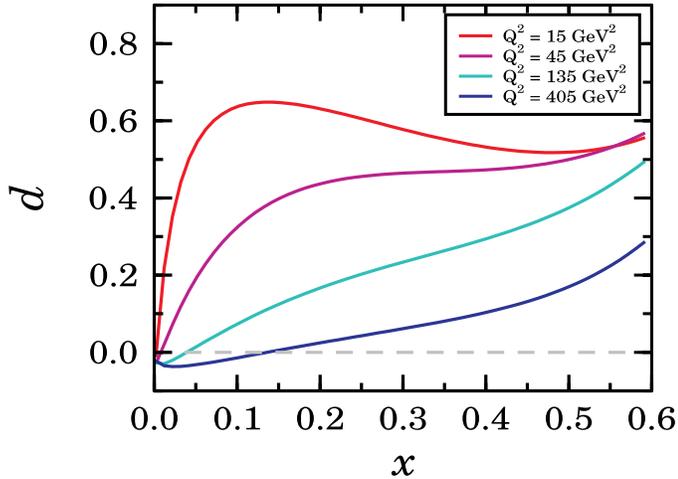}
\caption{Relative deviation of calculated cross section of DIS
off the proton from parton model,
$d=(\sigma_{ISI}-\sigma_{LO})/(\sigma_{ISI}+\sigma_{LO})$, for
different $Q^2$.} \label{dis2}
\end{figure}

The found effect of the parton virtuality in DIS is
$Q^2$-suppressed (figure~\ref{dis2}). For the most of the
experimentally investigated values of $Q^2$, the ambiguity in the
parton distribution function parameterizations due to the
renormalization scale uncertainty is of the same order as the ISI
effect in DIS. Thus, the value of the quark width in the nucleon
cannot be extracted from the DIS data. This is the result which
was expected by an analogy to nuclear physics, because the DIS
cross section is fully inclusive.

\begin{figure*}
\begin{center}
\resizebox{0.75\textwidth}{!}{%
  \includegraphics{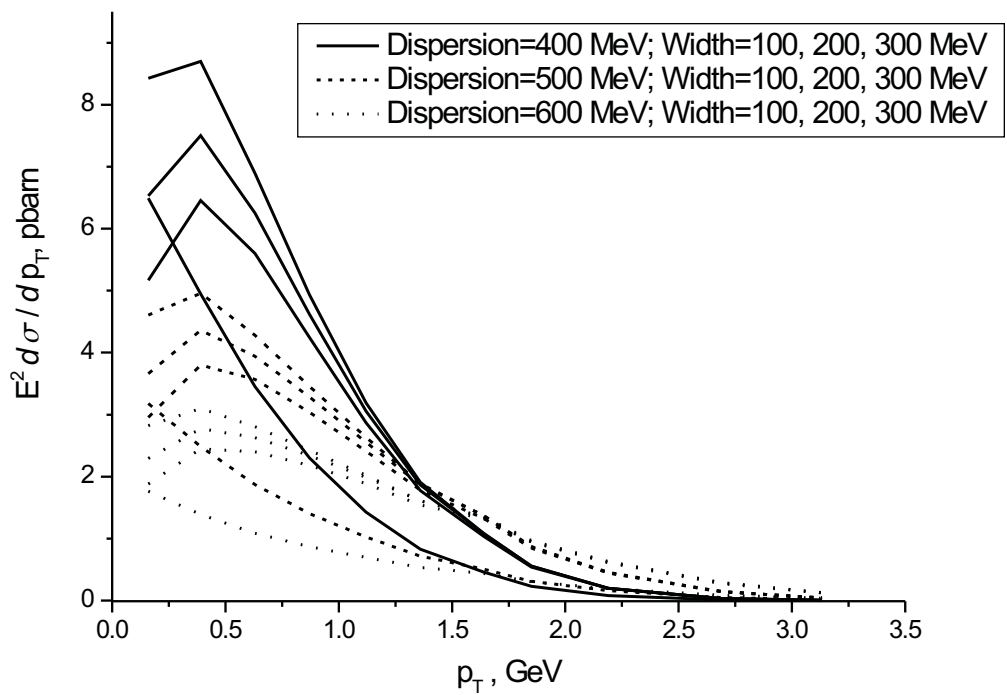}
} \caption{Calculated distribution of the Drell-Yan lepton pair's
transverse momentum for different values of the quark primordial
transverse momentum dispersion and spectral function width.}
\label{variab}
\end{center}
\end{figure*}
\begin{figure*}
\begin{center}
\resizebox{0.75\textwidth}{!}{%
  \includegraphics{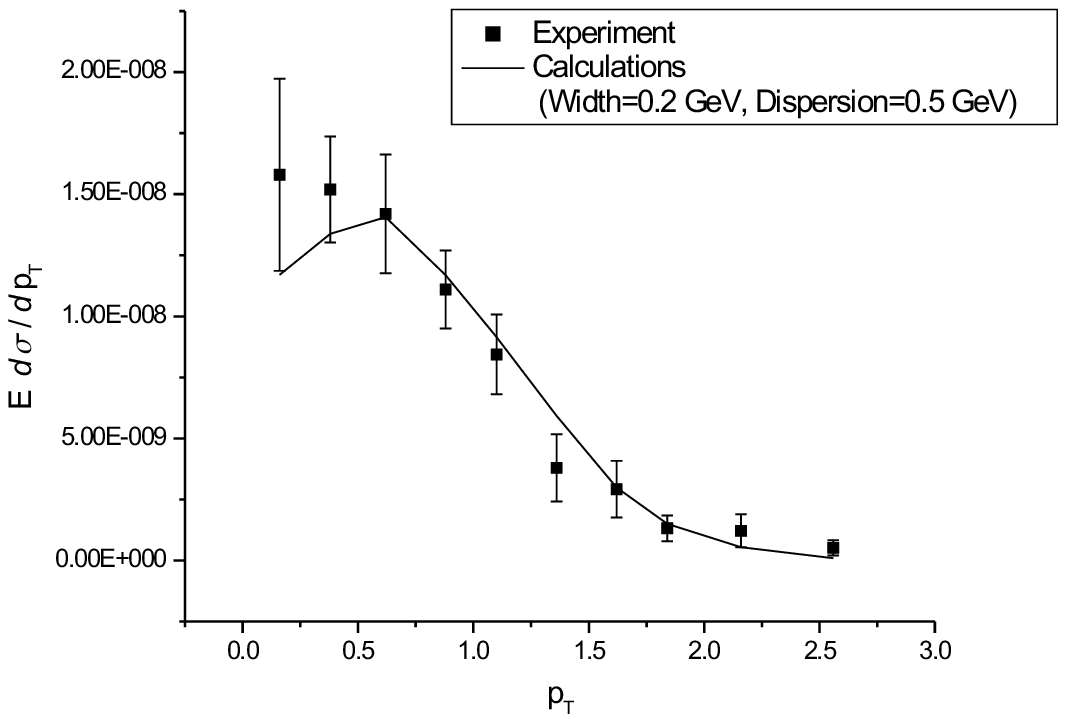}
} \caption{Calculation result, compared to the data of the
Fermilab experiment 866 for the continuum dimuon production in
800~GeV/c proton collision, $5\le M\le 7$~GeV, $-0.05\le x_F\le
0.2$. Only statistical errors shown.} \label{fit2}
\end{center}
\end{figure*}
\begin{figure*}
\begin{center}
\resizebox{0.75\textwidth}{!}{%
  \includegraphics{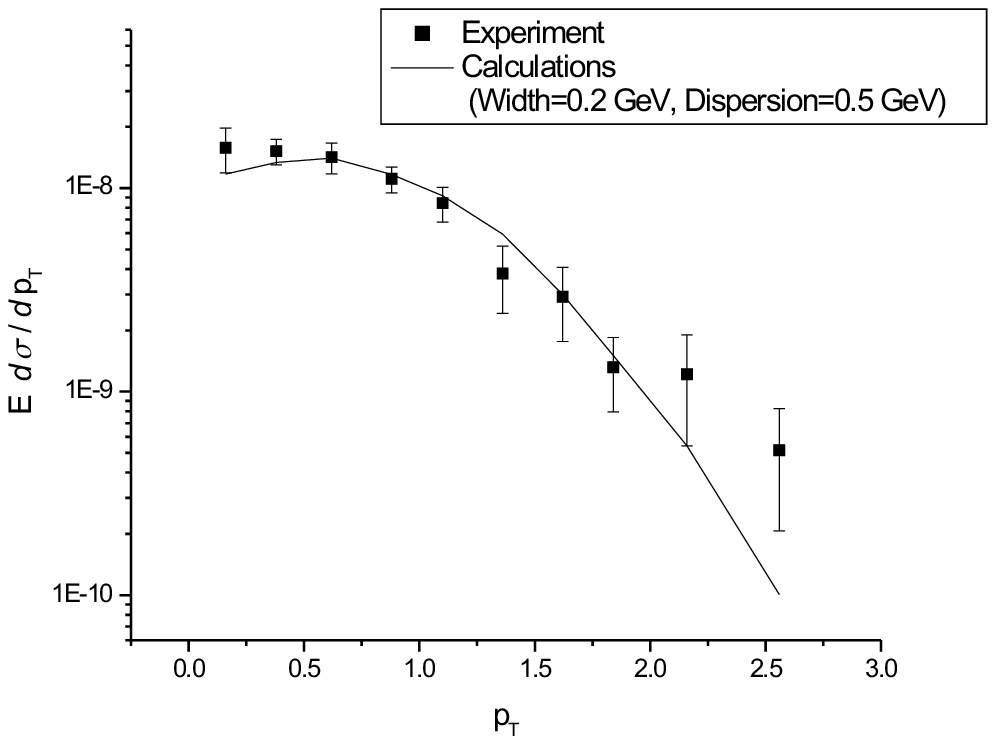}
} \caption{Calculation result, compared to the data of the
Fermilab experiment 866 for the continuum dimuon production in
800~GeV/c proton collision, $5\le M\le 7$~GeV, $-0.05\le x_F\le
0.2$. Only statistical errors shown. } \label{fit1}
\end{center}
\end{figure*}

In contrast, for such a semi-exclusive observable as the
transverse momentum
distribution of the Drell-Yan lepton pair, we have found a
substantial dependence on both the dispersion of the quark
primordial transverse momentum and the spectral function width
(figure \ref{variab}). The LO QCD prediction for this distribution
is the $\delta$-function around~$0$, while the experimentally 
measured distribution is rather broad.

On the figures \ref{fit1} and \ref{fit2} an example of our
description of the experimental data is presented. In this data
set, the mass of the Drell-Yan pair is around $7$~GeV and
$0<x_F<0.2$. The optimal parameters for these values of $M_{DY}$
and $x_F$ are 500~MeV for the dispersion and 200~MeV for the
width. The data are reproduced very well over the two orders of
magnitude. The slight underestimation at $p_T>2.5$~GeV is caused
by the considerable contribution of the gluon Compton scattering
process to the measurable cross section at these high $p_T$. 
Note
that the convex shape 
of the
 
distribution at 
small $p_T$ 
cannot be 
described
 
without the inclusion of the off-shell effects
(figure~\ref{variab}).
In addition, by other models, neglecting off-shellness,
the magnitude of the cross section is not correctly obtained and
an additional overall K-factor is used. Our calculations yield not
only the experimentally measured form of the cross section but
also its amplitude without any K-factor.

\section{Summary and outlook}

\label{summary}

We 
developed a formalism 
to study the quark 
and gluon 
structure of
hadrons going further than the well known picture of collinear
non-interacting partons and applied it to calculate cross sections
of several high-energy processes. In this paper, the deep
inelastic scattering and Drell-Yan pair production are considered.
We explicitly took into account the quark initial state
interaction, missed in the standard perturbative consideration, by
dressing the quark lines with spectral functions and using the
method of generalized factorization.

There was discovered a substantial contribution of the quark
off-shellness to the transverse momentum distribution of high-mass
virtual photons produced in high-energy hadron-hadron collisions.
The quark width in proton was estimated from comparison to the
resent data of the experiment E866 at Fermilab. 
For the mass of the Drell-Yan pair around $7$~GeV and
 
$0<x_F<0.2$, the quark width is 200 MeV.
More details about
the method and the results can be found in the long write-up by
the same authors~\cite{us}.


\begin{thebibliography}{}
%
\bibitem{kt}
X.-N.~Wang., Phys.Rev. \textbf{C61}, (2000) 064910;
Y.~Zhang,G.~Fai,G.~Papp,G.~Barnafoldi, and P.~Levai, Phys.Rev.
\textbf{C65} (2002) 034903
\bibitem{grv}
M.~Glueck,E.~Reya,A.~Vogt,Eur.Phys.J. \textbf{C5}, (1998) 461
\bibitem{us}
O.~Linnyk,S.~Leupold,U.~Mosel, in preparation
\bibitem{exp}
J.C.~Webb,PhD theses, hep-ex/0301031
\end{thebibliography}
\end{document}